# Exact solution of the Schrödinger equation for a short-range exponential potential with inverse square root singularity

**A.M. Ishkhanyan**[1,2]

[1]Institute for Physical Research, NAS of Armenia, Ashtarak 0203, Armenia
[2]Institute of Physics and Technology, National Research Tomsk Polytechnic University, Tomsk 634050, Russia

We introduce an exactly integrable singular potential for which the solution of the one-dimensional stationary Schrödinger equation is written through irreducible linear combinations of the Gauss hypergeometric functions. The potential, which belongs to a general Heun family, is a short-range one that behaves as the inverse square root in the vicinity of the origin and vanishes exponentially at the infinity. We derive the exact spectrum equation for the energy and discuss the bound states supported by the potential.

## 1. Introduction

The irreducible multi-term solutions of non-relativistic and relativistic wave equations which have been first noticed in connection with piecewise continuous potentials [1-3] and further in the context of super-symmetric quantum mechanics [4,5] have gained increasing attention during the last few years due to several new exactly [6-9] or conditionally [10,11] integrable as well as certain non-analytic [12-14] potentials recently reported. A particular subset of such solutions also recently noticed involves the rationally extended potentials [15-17] for which the solution is written in terms of exceptional orthogonal polynomials [18]. The potentials allowing irreducible multi-term solutions form a separate class of potentials which is out of the Natanzon family [19] to which the five classical exactly integrable [20-24] as well as the most discussed conditionally integrable [25-27] potentials belong.

In the present paper we introduce one more exactly integrable potential for which the solution of the one-dimensional stationary Schrödinger equation is written through irreducible linear combinations of the Gauss hypergeometric functions. As all novel exactly or conditionally exactly integrable potentials and many rationally extended ones, the potential belongs to the Heun class of potentials first discussed by Lemieux and Bose [28]. We have recently classified all Heun cases for which all the involved parameters can be varied independently and have shown that there exist in total 29 independent such families [29-31]. The potential we introduce belongs to the fifth general Heun family [31] which presents a generalization of the Eckart [23] and the third exactly solvable [9] hypergeometric potentials.

This is a singular potential which behaves as the inverse square root in the vicinity of the origin and vanishes exponentially at infinity. Hence, it is a short-range potential that supports only a finite number of bound states. We present the general solution of the problem conveniently written in terms of the Clausen generalized hypergeometric functions [32,33]. This solution has an alternative representation given through irreducible linear combinations of the Gauss ordinary hypergeometric functions. We derive the exact equation for the energy spectrum and construct the zero-energy solution of the Schrödinger equation the zeros of which give the exact number of bound states.

## 2. The potential and the solution

The potential we consider is

$$V = V_0 + \frac{V_1}{\sqrt{1-e^{-x/\sigma}}}.  \quad (1)$$

For a positive $\sigma > 0$ this is a potential well defined on the positive semi-axis, and for $V_1 = -V_0$ it vanishes at infinity.

Following the lines of [31], we apply the transformation of the dependent and independent variables

$$\psi = \varphi(z)\, u(z), \quad z = z(x)  \quad (2)$$

with

$$\varphi = (z+1)^{\alpha_1} (z-1)^{\alpha_2}  \quad (3)$$

and

$$z = \sqrt{1-e^{-x/\sigma}}  \quad (4)$$

to reduce the one-dimensional Schrödinger equation for potential (1)

$$\frac{d^2\psi}{dx^2} + \frac{2m}{\hbar^2}(E - V(x))\psi = 0  \quad (5)$$

to the general Heun equation [34]

$$\frac{d^2u}{dz^2} + \left(\frac{\gamma}{z} + \frac{\delta}{z-1} + \frac{\varepsilon}{z+1}\right)\frac{du}{dz} + \frac{\alpha\beta z - q}{z(z-1)(z+1)} u = 0.  \quad (6)$$

This reduction is verified by checking that transformation (2) results in the equation

$$u_{zz} + \left(2\frac{\varphi_z}{\varphi} + \frac{\rho_z}{\rho}\right) u_z + \left(\frac{\varphi_{zz}}{\varphi} + \frac{\rho_z}{\rho}\frac{\varphi_z}{\varphi} + \frac{2m}{\hbar^2}\frac{E-V(z)}{\rho^2}\right) u = 0  \quad (7)$$

with $\rho = dz/dx$ and noting that for the coordinate transformation (4) we have

$$\rho = \frac{dz}{dx} = -\frac{(z+1)(z-1)}{2\sigma z}.  \quad (8)$$

It is then readily checked, by substitution of equations (3) and (8) into equation (7), that the parameters of the Heun equation are given as

$$(\gamma, \delta, \varepsilon) = (-1, 1 + 2\alpha_2, 1 + 2\alpha_1),\tag{9}$$

$$\alpha, \beta = \alpha_1 + \alpha_2 \pm \sqrt{\frac{8m\sigma^2}{\hbar^2}(-E + V_0)}, \quad q = -\alpha_1 + \alpha_2.\tag{10}$$

$$\alpha_1 = \pm\sqrt{\frac{2m\sigma^2}{\hbar^2}(-E + V_0 - V_1)}, \quad \alpha_2 = \pm\sqrt{\frac{2m\sigma^2}{\hbar^2}(-E + V_0 + V_1)}.\tag{11}$$

We note that $\gamma = -1$ so that the characteristic exponents $\mu_{1,2} = 0, 1 - \gamma$ [34] of singularity $z = 0$ are 0 and 2. Since the difference between the exponents is integer, the Frobenius series $u = z^\mu \sum_{n=0}^{\infty} c_n z^n$ are generally expected to produce only one consistent solution - that with the greater exponent $\mu_2 = 2$. The other independent solution of the Heun equation is generally expected to involve a logarithmic term. However, it is checked that for the particular case at hand the singularity is in fact *apparent* (simple) [35] which means that the logarithmic term disappears and the general solution is analytic in the singularity.

Indeed, let us consider the expansion of the solution of the Heun equation as a Taylor series $u = \sum_{n=0}^{\infty} c_n z^n$ with $c_0 \neq 0$. Substituting this into the Heun equation (6) with $\gamma = -1$ and successively calculating the coefficients $c_n$, for $c_2$ we face a division by zero unless

$$q^2 + q(\varepsilon - 1 + a(\delta - 1)) + a\alpha\beta = 0,\tag{12}$$

where $a$ is the third regular singularity of the Heun equation which, in our case, is $a = -1$. This is the condition for the singularity $z = 0$ to be apparent. It is immediately checked that for the parameters given by equations (9)-(11) this condition is satisfied.

Now we look for hypergeometric representations of the Taylor series solutions. We first check that the Heun-to-hypergeometric reductions through the common one-term ansatz involving a single Gauss hypergeometric function $_2F_1$ (see, e.g., [36,37]) do not apply. Helpfully, Letessier and co-authors have noticed [38] (see also [39]) that for a root of equation (12) the solution of the Heun equation (6) has a representation in terms of the Clausen generalized hypergeometric function $_3F_2$ [32,33]. This representation reads

$$u_1 = {}_3F_2\left(\alpha, \beta, 1 + \frac{\alpha\beta}{q}; \frac{\alpha\beta}{q}, \varepsilon; \frac{z+1}{2}\right).\tag{13}$$

The second independent solution of the Heun equation is checked to be

$$u_2 = {}_3F_2\left(\alpha, \beta, 1-\frac{\alpha\beta}{q}; -\frac{\alpha\beta}{q}, \delta; \frac{1-z}{2}\right). \tag{14}$$

Thus, the general solution of the Schrödinger equation for potential (1) is given as

$$\psi = (z+1)^{\alpha_1}(z-1)^{\alpha_2}\left(c_1 u_1 + c_2 u_2\right). \tag{15}$$

We conclude this section by noting that the involved Clausen generalized hypergeometric functions have alternative representation in terms of the familiar Gauss ordinary hypergeometric functions. This representation is readily derived by the series expansion of the Heun function in terms of the Gauss functions [40-42]. The result derived through the termination of such a series on the second term, for a root of equation (12), is a generally irreducible linear combination of two Gauss functions:

$$u_1 = {}_2F_1\left(\alpha, \beta; \varepsilon; \frac{1+z}{2}\right) - \frac{2\alpha_1}{\alpha_1+\alpha_2} \cdot {}_2F_1\left(\alpha, \beta; \varepsilon-1; \frac{1+z}{2}\right). \tag{16}$$

This solution applies for any real or complex set of the involved parameters with the proviso that $\varepsilon$ is not unity, zero, or a negative integer. Comparing the corresponding terms in the power-series expansions, it is checked that this result differs from the Clausen function (13) by only the non-essential constant factor $(\alpha_2 - \alpha_1)/(\alpha_2 + \alpha_1)$.

## 3. Bound states

In order for the potential (1) to vanish at infinity we put $V_1 = -V_0$ so that

$$V = V_0 - \frac{V_0}{\sqrt{1-e^{-x/\sigma}}}. \tag{17}$$

The shape of the potential is shown in Fig. 1. The behavior of the potential in the vicinity of the origin resembles that of the inverse square root potential [6]:

$$V\big|_{x\to 0} \sim -\frac{V_0}{\sqrt{x/\sigma}}, \tag{18}$$

however, this is a short-range well because it vanishes at infinity exponentially:

$$V\big|_{x\to+\infty} \sim -\frac{V_0}{2}e^{-\frac{x}{\sigma}}. \tag{19}$$

We note that these asymptotes are the same as those for the Lambert-W singular potential [8]. The integral of the function $xV(x)$ over the positive semi-axis is finite; hence, the potential supports only a finite number of bound states [43].

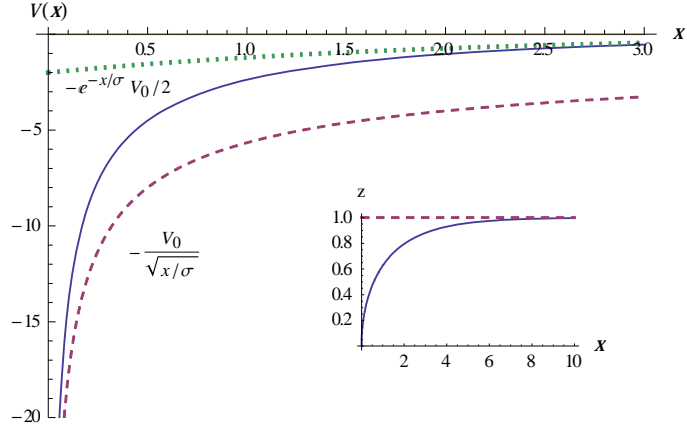

Fig.1. Potential (17) for $m, \hbar, V_0, \sigma = 1, 1, 4, 2$ (solid line). The dashed line presents the asymptote (18) for $x \to 0$ and the dotted line stands for the exponential asymptote (19) for $x \to \infty$. The inset presents the coordinate transformation $z(x)$.

The bound states are derived by demanding the wave function to vanish both at the origin and at infinity (see the discussion in [44]). For the plus signs for $\alpha_{1,2}$ in equation (11), that we choose for convenience, the second of these requirements results in $c_1 = 0$ for solution (15). The first condition then immediately produces the exact equation for the energy spectrum

$$_3F_2\left(\alpha, \beta, 1 - \frac{\alpha\beta}{q}; -\frac{\alpha\beta}{q}, \delta; \frac{1}{2}\right) = 0. \tag{20}$$

This equation can be alternatively presented in terms of Gauss functions as

$$S(E) \equiv 1 - \frac{\alpha\beta + 2\alpha_2 q}{2\alpha_2 q} \frac{_2F_1(\alpha, \beta, 1 + 2\alpha_2; 1/2)}{_2F_1(\alpha, \beta, 2\alpha_2; 1/2)} = 0. \tag{21}$$

The graphical representation of the $S(E)$ function is shown in Fig. 2.

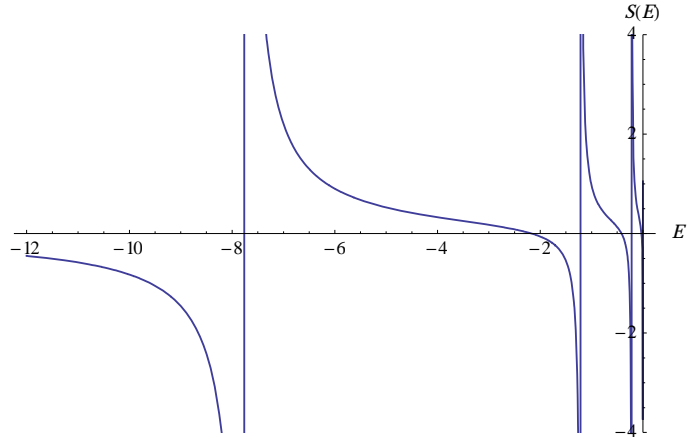

Fig.2. Graphical representation of the spectrum equation (21) for $m, \hbar, V_0, \sigma = 1, 1, 4, 2$.

With the spectrum equation (20) or (21), and recalling that $V_1 = -V_0$, the bound state wave functions are explicitly written in terms of the Gauss functions as

$$\psi_B = (z+1)^{\alpha_1}(z-1)^{\alpha_2}\left({}_2F_1\left(\alpha,\beta;1+2\alpha_2;\frac{1-z}{2}\right) - \frac{2\alpha_2 q}{\alpha\beta + 2\alpha_2 q}\,{}_2F_1\left(\alpha,\beta;2\alpha_2;\frac{1-z}{2}\right)\right), \quad (22)$$

where $z = \sqrt{1-e^{-x/\sigma}}$, and

$$\alpha,\beta = \alpha_1 + \alpha_2 \pm \sqrt{\frac{8m\sigma^2}{\hbar^2}(-E+V_0)}, \quad q = -\alpha_1 + \alpha_2, \quad (23)$$

$$\alpha_1 = \sqrt{\frac{2m\sigma^2}{\hbar^2}(-E+2V_0)}, \quad \alpha_2 = \sqrt{\frac{-2m\sigma^2 E}{\hbar^2}}. \quad (24)$$

The normalized wave functions and energy levels are shown in Fig. 3.

The number of bound states $n$ can be rather accurately estimated by checking the integral by Calogero [45]

$$n \leq I_C = \frac{2/\pi}{\hbar/\sqrt{2m}} \int_0^\infty \sqrt{-V(x)}\,dx, \quad (25)$$

which is specialized for everywhere monotonically non-decreasing attractive potentials. As it can be seen, this estimate states that for strongly attractive potentials the number of bound states increases as the square root of the strength of the potential. The asymptotic result by Chadan further tunes the upper limit of the number of bound states to a half of Calogero's integral [46]. For potential (17) the result is

$$n \leq \frac{I_C}{2} = 2(\sqrt{2}-1)\sqrt{\frac{m\sigma^2 V_0}{\hbar^2}}. \quad (26)$$

For the parameters used in Fig. 2 we have $n \leq 3.31$, that is the number of bound states is three. The numerical spectra for $V_0 = 4, 6, 10, 15$ ($m, \hbar, \sigma = 1, 1, 2$) are presented in Table 1.

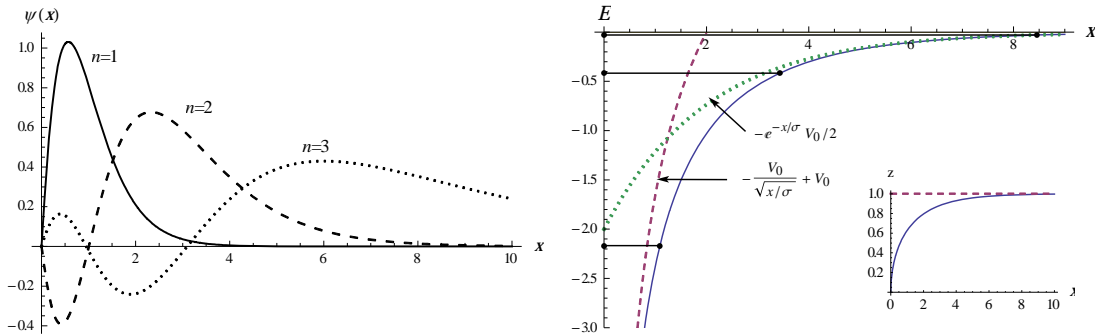

Fig. 3. The first three normalized wave functions and the energy levels for $m, \hbar, V_0, \sigma = 1, 1, 4, 2$.

| $V_0 = 4$ | $V_0 = 6$ | $V_0 = 10$ | $V_0 = 15$ |
|---|---|---|---|
| $-0.02946950533$ | $-0.00552717659$ | $-0.02016478163$ | $-0.04351527517$ |
| $-0.4166327432$ | $-0.2384895435$ | $-0.3060888017$ | $-0.3819693937$ |
| $-2.168051138$ | $-1.120668116$ | $-1.146332147$ | $-1.227239451$ |
| | $-4.338949251$ | $-3.249038689$ | $-3.007062709$ |
| | | $-9.973668269$ | $-6.947785749$ |
| | | | $-18.84341927$ |

Table 1. Numerical spectra $E_n$ for $V_0 = 4, 6, 10, 15$ ($m, \hbar, \sigma = 1, 1, 2$)

The exact number of bound states is known to be equal to the number of zeros (not counting $x = 0$) of the zero-energy solution, which vanishes at the origin [43]. Since for $E = 0$ the exponent $\alpha_2$ becomes zero and the lower parameter $\delta$ of the second independent solution (14) becomes unity, in constructing this solution one should be careful with the sign of the exponent $\alpha_1$. It is checked that the consistent sign for the second independent solution is the minus. Hence, the general solution of the Schrödinger equation for $E = 0$ reads

$$\psi_{E=0} = c_1 (z+1)^{\alpha_1} {}_3F_2\left(\sqrt{2}\alpha_1 + \alpha_1, -\sqrt{2}\alpha_1 + \alpha_1, 1+\alpha_1; \alpha_1, 1+2\alpha_1; \frac{z+1}{2}\right) + $$
$$c_2 (z+1)^{-\alpha_1} {}_3F_2\left(\sqrt{2}\alpha_1 - \alpha_1, -\sqrt{2}\alpha_1 - \alpha_1, 1+\alpha_1; \alpha_1, 1; \frac{1-z}{2}\right), \qquad (27)$$

where $\alpha_1 = 2\sqrt{m\sigma^2 V_0 / \hbar^2}$. The condition $\psi_{E=0}(0) = 0$ then produces a linear relation between $c_1$ and $c_2$, which finalizes the construction of the needed solution.

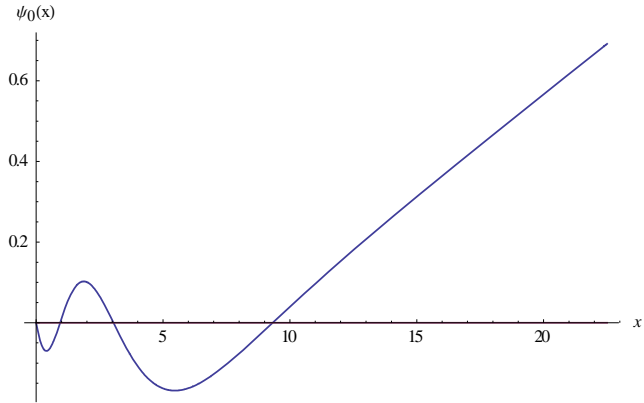

Fig.4. The zero-energy solution for $m, \hbar, V_0, \sigma = 1, 1, 4, 2$.

This solution for the parameters applied in Fig. 2 is shown in Fig. 4. It can be seen that the number of bound states is indeed three. A last remark that may be useful for applications of the zero-energy solution itself, e.g., in constructing super-symmetrically extended potentials or designing optical field configurations for optimal control purposes, is that the solution diverges at the infinity with a logarithmic asymptote.

## 5. Scattering states

As it was already mentioned above, the general solution (15) applies for any real or complex set of the involved parameters with the proviso that $\varepsilon$ is not unity, zero, or a negative integer. Consider then the scattering solution for which $E>0$. This time, one imposes the boundary condition of vanishing the wave-function in the origin and looks for an asymptote at the infinity as a combination of incoming and outgoing fluxes.

The boundary condition for the origin imposes

$$c_2 = -\frac{{}_3F_2(\alpha,\beta,1+\alpha\beta/q;\alpha\beta/q,\varepsilon;1/2)}{{}_3F_2(\alpha,\beta,1-\alpha\beta/q;-\alpha\beta/q,\delta;1/2)}c_1. \tag{28}$$

Discussing now the behavior at infinity, we note that at $z\to 1$ the Clausen functions (13) and (14) have the asymptotes

$$u_1 \sim \frac{(q-\delta+1)\Gamma(1-\delta)\Gamma(\varepsilon)}{\Gamma(\alpha-\delta+2)\Gamma(\varepsilon-\alpha)} + \frac{2^{\delta-1}q\Gamma(\delta-1)\Gamma(\varepsilon)}{\Gamma(1+\alpha)\Gamma(1+\beta)}(1-z)^{1-\delta}, \quad u_2 \sim 1. \tag{29}$$

Accordingly, the large-distance behavior of the wave function is derived to be

$$\psi(x)\big|_{x\to\infty} \sim (-1)^{\alpha_2} 2^{\alpha_1-\alpha_2}\left(Ae^{-\frac{\alpha_2 x}{2}} + Be^{\frac{\alpha_2 x}{2}}\right)c_1, \tag{30}$$

where

$$A = \frac{c_2}{c_1} - \frac{(\alpha_1+\alpha_2)\Gamma(1+2\alpha_1)\Gamma(-2\alpha_2)}{\Gamma(1-\alpha+2\alpha_1)\Gamma(1+\alpha-2\alpha_2)}, \quad B = \frac{16^{\alpha_2}(\alpha_2-\alpha_1)\Gamma(1+2\alpha_1)\Gamma(2\alpha_2)}{\Gamma(1+\alpha)\Gamma(1-\alpha+2\alpha_1+2\alpha_2)}. \tag{31}$$

With this, for the phase shift $\delta_l$, which is defined by the representation

$$\psi(x) \sim e^{-ikx} + e^{2i\delta_l}e^{ikx}, \tag{32}$$

where $k$ is the wave-number

$$k = \alpha_2/(2i) = \sqrt{\frac{m\sigma^2 E}{2\hbar^2}}, \tag{33}$$

we get

$$\delta_l = \frac{i}{2}\ln\left(\frac{A}{B}\right). \tag{34}$$

This is a real function the form of which (with the convention $|\delta_l|<\pi/2$) is shown in Fig.5.

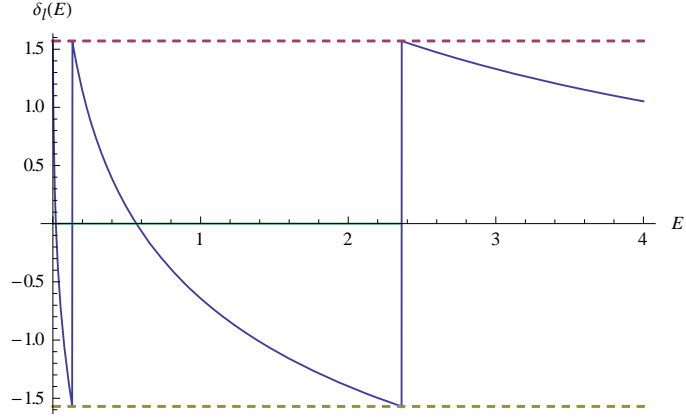

Fig.5. The phase shift $\delta_l$ for $m, \hbar, V_0, \sigma = 1, 1, 4, 2$. The dashed lines indicate $\delta_l = \pm \pi / 2$.

We observe that for some values of energy the phase shift shows a jump equal to $\pi$. This occurs if $\delta_l = \pm \pi / 2$ or, equivalently, if $A = -B$. For a fixed potential strength $V_1$, the number of jumps is finite. It is checked that this number is less by one than the number of bound-states supported by the potential.

## 6. Discussion

Thus, we have introduced a singular Schrödinger potential exactly integrable in terms of the Gauss hypergeometric functions. The potential presents a short-range bottomless well that behaves as the inverse square root in the vicinity of the origin and vanishes exponentially at the infinity. We note that these asymptotes are the same as those of the Lambert-W singular potential [8]. We have presented the general solution of the Schrödinger equation, derived the exact spectrum equation for the bound-state energies and constructed the bound-state wave functions. Discussing the number of the bound states supported by the potential, we have presented an accurate estimate and derived the zero-energy solution the number of zeros of which gives the exact number of the bound states.

The potential is obtained by reduction of the Schrödinger equation to the general Heun equation. This equation and its four confluent modifications have a remarkably large covering in contemporary physics and mathematics research [47]. Our potential is a four-parametric member of the fifth general Heun family [31] which presents an eight-parametric generalization of the classical five-parametric Eckart potential [23]. This family suggests several exactly or conditionally integrable sub-cases solvable in terms of the Gauss functions. In particular, our potential is the super-potential of the two conditionally solvable partner potentials by López-Ortega [11] which also belong to this family.

The solution of the Schrödinger equation for the potential we have presented is written through such fundamental solutions each of which presents an irreducible linear combination of two Gauss hypergeometric functions. A peculiarity of these fundamental solutions is that they have representations in terms of the Clausen generalized hypergeometric function $_3F_2$ [32,33]. We have checked that such a representation or a similar representation in terms of higher-order generalized hypergeometric functions $_pF_q$ is achieved also for all other known sub-potentials of the general Heun families allowing multi-term solutions as (irreducible) combinations of the Gauss hypergeometric functions. A common feature for all these representations is that at least an upper parameter exceeds a lower by unity as in our case of equations (13),(14) or (27). It is of interest if there exist solutions of the Schrödinger equation in terms of the generalized hypergeometric functions of different structure, i.e., in terms of generalized hypergeometric functions which are not reduced to the Gauss functions. Such solutions would provide a distinct class of solvable Schrödinger potentials.

### Acknowledgments

I thank the referee for very helpful comments and recommendations. This research has been conducted within the scope of the International Associated Laboratory IRMAS (CNRS-France & SCS-Armenia). The work has been supported by the Armenian State Committee of Science (SCS Grant No. 15T-1C323), Armenian National Science and Education Fund (ANSEF Grant No. PS-4986), and the project "Leading Russian Research Universities" (Grant No. FTI_24_2016 of the Tomsk Polytechnic University).